\documentclass[12pt]{article}%
\usepackage{amsmath}
\usepackage{amsfonts}
\usepackage{amssymb}
\usepackage{graphicx}
\usepackage{float}%
\begin{document}

\title{Spherical model with Dzyaloshinskii-Moriya interactions }
\author{William de Castilho and S. R. Salinas\\Instituto de F\'{\i}sica\\Universidade de S\~{a}o Paulo\\S\~{a}o Paulo, SP, Brazil}
\maketitle
\date{}

\begin{abstract}
We analyze the thermodynamic behavior of a ferromagnetic mean-spherical model
with three distinct spin components and the addition of Dzyaloshinkii-Moriya
interactions. Exact calculations are performed for classical and quantum
versions of this lattice model system. We show the onset of space modulated
structures at low temperatures.

\end{abstract}

\section{Introduction}

The spherical model of ferromagnetism, which was proposed and exactly solved
by Berlin and Kac about seventy years ago \cite{BerlinKac1952}
\cite{Joyce1972}, still remains an excellent laboratory to test some
thermodynamic properties of phase transitions and critical phenomena. We then
decided to investigate a ferromagnetic mean-spherical model, with three
distinct spin components and the addition of Dzyaloshinkii-Moriya (DM)
interactions \cite{Dzyaloshinskii1964} \cite{Moriya1960}. This model system is
still amenable to some exact calculations, both at the classical and at a
quantum level, which do indicate the presence of magnetically modulated
structures, which are the hallmark of the DM interactions \cite{Izyumov1984}
\cite{Togawa2016}.

Spherical models with several spin components, and with ferromagnetic
interactions, have been considered in the earlier literature, and have been
shown to lead to a simple factorization of the canonical partition function
\cite{BettoneyMazo}. A mean-spherical model has been recently used by Aqua and
Fisher \cite{AquaFisher} to account for some features of a lattice gas with
several components. We were then motivated to revisit these calculations for a
mean-spherical model, with the consideration of three spin components, and the
addition of DM interactions. The presence of extra couplings between two
different sets of degrees of freedom gives rise to a more involved and
interesting problem.

We consider the spin Hamiltonian
\begin{equation}
\mathcal{H}=-J\sum_{\left(  \overline{r},\overrightarrow{r}^{\prime}\right)
}\vec{S}_{\vec{r}}\cdot\vec{S}_{\vec{r}^{\prime}}-D\sum_{\vec{r}}\left(
\vec{S}_{\vec{r}}\times\vec{S}_{\vec{r}+\widehat{z}}\right)  \cdot\hat
{z},\label{hq}%
\end{equation}
where the classical spin variables $\vec{S}_{\vec{r}}=\left(
S_{\overrightarrow{r}}^{x},S_{\overrightarrow{r}}^{y},S_{\overrightarrow{r}%
}^{z}\right)  $ are three-component vectors on the sites $\overrightarrow{r}$
of a hypercubic lattice of $N$ sites, $J>0$ is a ferromagnetic exchange
interaction, the first sum is over nearest-neighbor pairs of lattice sites,
and $\widehat{z}$ is a unit vector along an axial direction. This is perhaps
the simplest spin Hamiltonian to represent a ferromagnetic model system with
the addition of monoaxial DM interactions.

The partition function of this classical mean-spherical model is usually
written as%
\begin{equation}
\Xi=\prod_{\vec{r}}\left[
{\displaystyle\int\limits_{-\infty}^{+\infty}}
dS_{\vec{r}}^{x}%
{\displaystyle\int\limits_{-\infty}^{+\infty}}
dS_{\vec{r}}^{y}%
{\displaystyle\int\limits_{-\infty}^{+\infty}}
dS_{\vec{r}}^{z}\right]  \exp\left[  \overline{\mathcal{H}}\right]
,\label{zetaq}%
\end{equation}
with%
\begin{equation}
\overline{\mathcal{H}}=-\beta\mathcal{H}-s_{1}\sum_{\vec{r}}(S_{\vec{r}}%
^{x})^{2}-s_{2}\sum_{\vec{r}}(S_{\vec{r}}^{y})^{2}-s_{3}\sum_{\vec{r}}%
(S_{\vec{r}}^{z})^{2},
\end{equation}
where $\beta=1/k_{B}T$ is the inverse of temperature, and $s_{1}$, $s_{2}$,
and $s_{3}$ are three spherical parameters. In this formulation, we have to
take into account three spherical constraints, which are given by%
\begin{equation}
\left\langle
{\displaystyle\sum\limits_{\overrightarrow{r}}}
(S_{\vec{r}}^{x})^{2}\right\rangle =-\frac{\partial}{\partial s_{1}}%
ln\Xi=N,\label{esf}%
\end{equation}
with similar equations for the $y$ and $z$ spin components,%
\begin{equation}
\left\langle
{\displaystyle\sum\limits_{\overrightarrow{r}}}
(S_{\vec{r}}^{y})^{2}\right\rangle =-\frac{\partial}{\partial s_{2}}%
ln\Xi=N,\quad\left\langle
{\displaystyle\sum\limits_{\overrightarrow{r}}}
(S_{\vec{r}}^{z})^{2}\right\rangle =-\frac{\partial}{\partial s_{3}}%
ln\Xi=N.\label{esf2}%
\end{equation}

In the first Section, we analyze the phase diagram of this classical system in
terms of temperature $T$ and a parameter $p=D/J$, which gauges the strength of
the chiral interactions. We perform exact calculations to show the existence
of a modulated structure along the $\widehat{z}$\ direction at sufficiently
low temperatures.

We then turn to the analysis of the quantum version of this mean-spherical
model with DM interactions. According to previous work for the mean-spherical
ferromagnet, a quantum version may be obtained by resorting to a standard
canonical quantization procedure \cite{Vojta1996} \cite{Oliveira2006}
\cite{Bienzobaz2012} \cite{Wald2015}. The problem is then formulated in terms
of a set of coupled boson operators, which are duly diagonalized by known
techniques of second quantization \cite{Bogoliubov1982}. Again, we show the
persistence of spacial modulated structures in the low-temperature region of
the phase diagram.

\section{Mean-spherical model with DM interactions}

The partition function of the three-component mean-spherical model with DM
interactions is given by equation (\ref{zetaq}) supplemented by the spherical
constraints, eq. (\ref{esf}) and (\ref{esf2}). We now introduce periodic
boundary conditions, and write a Fourier representation,
\begin{equation}
S_{\overrightarrow{r}}^{\nu}=\frac{1}{\sqrt{N}}%
{\displaystyle\sum\limits_{\overrightarrow{q}}}
\sigma_{\overrightarrow{q}}^{\nu}\,\exp\left(  i\overrightarrow{q}%
\cdot\overrightarrow{r}\right)  ,
\end{equation}
where $\nu=x,y,z$, and the sum is over a symmetric Brillouin zone. The problem
is then reduced to the diagonalization of a quadratic form,%
\[
\overline{\mathcal{H}}=\beta J\sum_{\vec{q}}\left(  \cos q_{x}+\cos q_{y}+\cos
q_{z}\right)  \left(  \sigma_{\vec{q}}^{x}\sigma_{-\vec{q}}^{x}+\sigma
_{\vec{q}}^{y}\sigma_{-\vec{q}}^{y}+\sigma_{\vec{q}}^{z}\sigma_{-\vec{q}}%
^{z}\right)  +
\]%
\begin{equation}
-2\beta Di\sum_{\vec{q}}\left(  \sin q_{z}\right)  \,\sigma_{\vec{q}}%
^{x}\sigma_{-\vec{q}}^{y}-s_{1}\sum_{\vec{q}}(\sigma_{\vec{q}}^{x}%
\sigma_{-\vec{q}}^{x})-s_{2}\sum_{\vec{q}}(\sigma_{\vec{q}}^{y}\sigma
_{-\vec{q}}^{y})-s_{3}\sum_{\vec{q}}(\sigma_{\vec{q}}^{z}\sigma_{-\vec{q}}%
^{z}).
\end{equation}

With the introduction of a standard orthogonal transformation,%
\begin{equation}
\sigma_{\vec{q}}^{\alpha}=\frac{1}{\sqrt{2}}\left(  R_{q}^{\alpha}%
+iI_{q}^{\alpha}\right)  ,\qquad q\neq0;\qquad\sigma_{0}^{\alpha}%
=R_{0}^{\alpha},
\end{equation}
where%
\begin{equation}
R_{\vec{q}}^{\alpha}=R_{-\vec{q}}^{\alpha},\qquad I_{q}^{\alpha}%
=-I_{-q}^{\alpha},
\end{equation}
with $\alpha=x,y,z$, we write the quadratic expression%
\[
\overline{\mathcal{H}}=\beta J\sum_{\vec{q}\geq0}\left(  \cos q_{x}+\cos
q_{y}+\cos q_{z}\right)  \,[(R_{\vec{q}}^{x})^{2}+(I_{\vec{q}}^{x}%
)^{2}+(R_{\vec{q}}^{y})^{2}+(I_{\vec{q}}^{y})^{2}+(R_{\vec{q}}^{z}%
)^{2}+(I_{\vec{q}}^{z})^{2}]+
\]%
\[
+2\beta D\sum_{\vec{q}\geq0}(\sin q_{z})\,[R_{\vec{q}}^{x}I_{\vec{q}}%
^{y}-R_{\vec{q}}^{y}I_{\vec{q}}^{x}]-
\]%
\begin{equation}
-s_{1}\sum_{\vec{q}\geq0}[(R_{\vec{q}}^{x})^{2}+(I_{\vec{q}}^{x})^{2}%
]-s_{2}\sum_{\vec{q}\geq0}[(R_{\vec{q}}^{y})^{2}+(I_{\vec{q}}^{y})^{2}%
]-s_{3}\sum_{\vec{q}\geq0}[(R_{\vec{q}}^{z})^{2}+(I_{\vec{q}}^{z})^{2}].
\end{equation}
Due to the DM couplings, this expression still requires a further
transformation to be written in a diagonal form.

\subsection{Critical ferromagnetic border}

In order to carry out the diagonalization of this problem, we consider two
distinct quadratic forms. One of these forms, which involves just the $z$
components of the spin variables, is already diagonal,%
\begin{equation}
Q_{1}=\left[  \beta J(\cos q_{x}+\cos q_{y}+\cos q_{z})-s_{3}\right]
[(R_{\vec{q}}^{z})^{2}+(I_{\vec{q}}^{z})^{2}].\label{Q1}%
\end{equation}
Introducing the spherical potential $\mu_{3}=s_{3}/\beta$, we require the
inequality%
\begin{equation}
\mu_{3}>\max\limits_{\overrightarrow{q}}\left[  J(\cos q_{x}+\cos q_{y}+\cos
q_{z})\right]  ,
\end{equation}
which is the usual condition associated with the existence of a ferromagnetic
phase transition in the ferromagnetic spherical model.

The quadratic form (\ref{Q1}) contributes to the partition function with a
term that depends on $s_{3}$. We then write this contribution,
\begin{equation}
\ln\Xi=...-\frac{1}{2}\frac{N}{\left(  2\pi\right)  ^{3}}\int d^{3}%
\overrightarrow{q}\ln\left[  s_{3}-\beta J(\cos q_{x}+\cos q_{y}+\cos
q_{z})\right]  +.....
\end{equation}
Using equation (\ref{esf}) for the spherical constraint, and taking the
maximum value of the spherical potential $\mu_{3}$, we obtain an integral
expression for the ferromagnetic critical border,%
\begin{equation}
J\beta_{c}=\frac{1}{2}\frac{1}{\left(  2\pi\right)  ^{3}}\int d^{3}%
\overrightarrow{q}\frac{1}{3-\left(  \cos q_{x}+\cos q_{y}+\cos q_{z}\right)
},\label{fborder}%
\end{equation}
which can be compared with the well-known result for the simple spherical
ferromagnet on a cubic lattice \cite{Joyce1972}.

\subsection{Critical border of a modulated structure}

We now simplify the notation and write the remaining, non diagonal, part of
the quadratic form,%
\begin{equation}
Q_{2}=A\left(  x_{1}^{2}+y_{1}^{2}\right)  +C\left(  x_{2}^{2}+y_{2}%
^{2}\right)  +B\left(  x_{1}y_{2}-y_{1}x_{2}\right)  ,
\end{equation}
with%
\begin{equation}
A=\beta J(\cos q_{x}+\cos q_{y}+\cos q_{z})-s_{1},
\end{equation}%
\begin{equation}
B=2\beta D\sin q_{z},
\end{equation}
and%
\begin{equation}
C=\beta J(\cos q_{x}+\cos q_{y}+\cos q_{z})-s_{2},
\end{equation}
in which $x_{1}=R_{q}^{x}$, $y_{1}=I_{q}^{x}$, $x_{2}=R_{q}^{y}$, $y_{2}%
=I_{q}^{y}$. In order to analyze this quadratic form, it is convenient to
introduce a $4\times4$ matrix,%
\begin{equation}
M=\left(
\begin{array}
[c]{cccc}%
A & 0 & 0 & \frac{1}{2}B\\
0 & A & -\frac{1}{2}B & 0\\
0 & -\frac{1}{2}B & C & 0\\
\frac{1}{2}B & 0 & 0 & C
\end{array}
\right)  ,
\end{equation}
which is symmetric, with real elements, and can be easily diagonalized. It is
straightforward to write the double-degenerate eigenvalues of this matrix,
\begin{equation}
\Lambda_{1,2}=\beta J(\cos q_{x}+\cos q_{y}+\cos q_{z})-\frac{1}{2}\left(
s_{1}+s_{2}\right)  \pm\frac{1}{2}\left[  \left(  s_{1}-s_{2}\right)
^{2}+4\beta^{2}D^{2}\sin^{2}q_{z}\right]  ^{1/2}.
\end{equation}

There are several consequences of the form of this expression for the
eigenvalues. Taking into account the symmetry between the exchange of
variables $s_{1}$ and $s_{2}$, and the form of the spherical constraints,
given by eqs. (\ref{esf}) and (\ref{esf2}), we can always make%
\begin{equation}
s_{1}=s_{2}=s=\beta\mu,
\end{equation}
and consider the much simpler expression%
\begin{equation}
\Lambda_{1,2}=\beta J(cosq_{x}+cosq_{y}+cosq_{z})-s\pm\beta D\sin q_{z},
\end{equation}
from which we obtain a maximum limit for the spherical potential,
\begin{equation}
\mu>\max\limits_{\overrightarrow{q}}\left\{  J(\cos q_{x}+\cos q_{y}+\cos
q_{z})\pm D\sin q_{z}\right\}  ,
\end{equation}
which can also be written as%
\begin{equation}
\mu>\max\limits_{q_{z}}\left\{  J(2+\cos q_{z})\pm D\sin q_{z}\right\}
.\label{mu1}%
\end{equation}
Therefore, we have a wave solution for the magnetization along the axis of
anisotropy, with a wave number $q_{z}$ given by
\begin{equation}
\tan q_{z}=\pm\frac{D}{J},\label{mu2}%
\end{equation}
which is a characteristic result associated with monoaxial DM interactions.

The free energy of this system contains terms of the form%
\[
\ln\Xi=...-\frac{1}{4}\frac{N}{\left(  2\pi\right)  ^{3}}\int d^{3}%
\overrightarrow{q}\ln\left[  s-\beta J(\cos q_{x}+\cos q_{y}+\cos q_{z})+\beta
D\sin q_{z}\right]  +.....
\]%
\begin{equation}
-\frac{1}{4}\frac{N}{\left(  2\pi\right)  ^{3}}\int d^{3}\overrightarrow{q}%
\ln\left[  s-\beta J(\cos q_{x}+\cos q_{y}+\cos q_{z})-\beta D\sin
q_{z}\right]  +....,
\end{equation}
from which we have%
\[
\beta=\frac{1}{4}\frac{1}{\left(  2\pi\right)  ^{3}}\int d^{3}\overrightarrow
{q}\,[\frac{1}{\mu-J(\cos q_{x}+\cos q_{y}+\cos q_{z})+D\sin q_{z}}+
\]%
\begin{equation}
+\frac{1}{\mu-J(\cos q_{x}+\cos q_{y}+\cos q_{z})-D\sin q_{z}}].
\end{equation}
The critical border of the modulated transition is given by this expression,
with the largest values of the spherical potential, according to equations
(\ref{mu1}) and (\ref{mu2}). In Figure 1, we draw this border in terms of
temperature $T$ and the parameter of chiral competition, $p=D/J$. This is a
typical result for a simple ferromagnetic system with monoaxial DM
interactions \cite{Bak1980}.%

\begin{figure}[H]
    \centering
    \includegraphics[width=0.8\textwidth, height=0.8\textwidth]{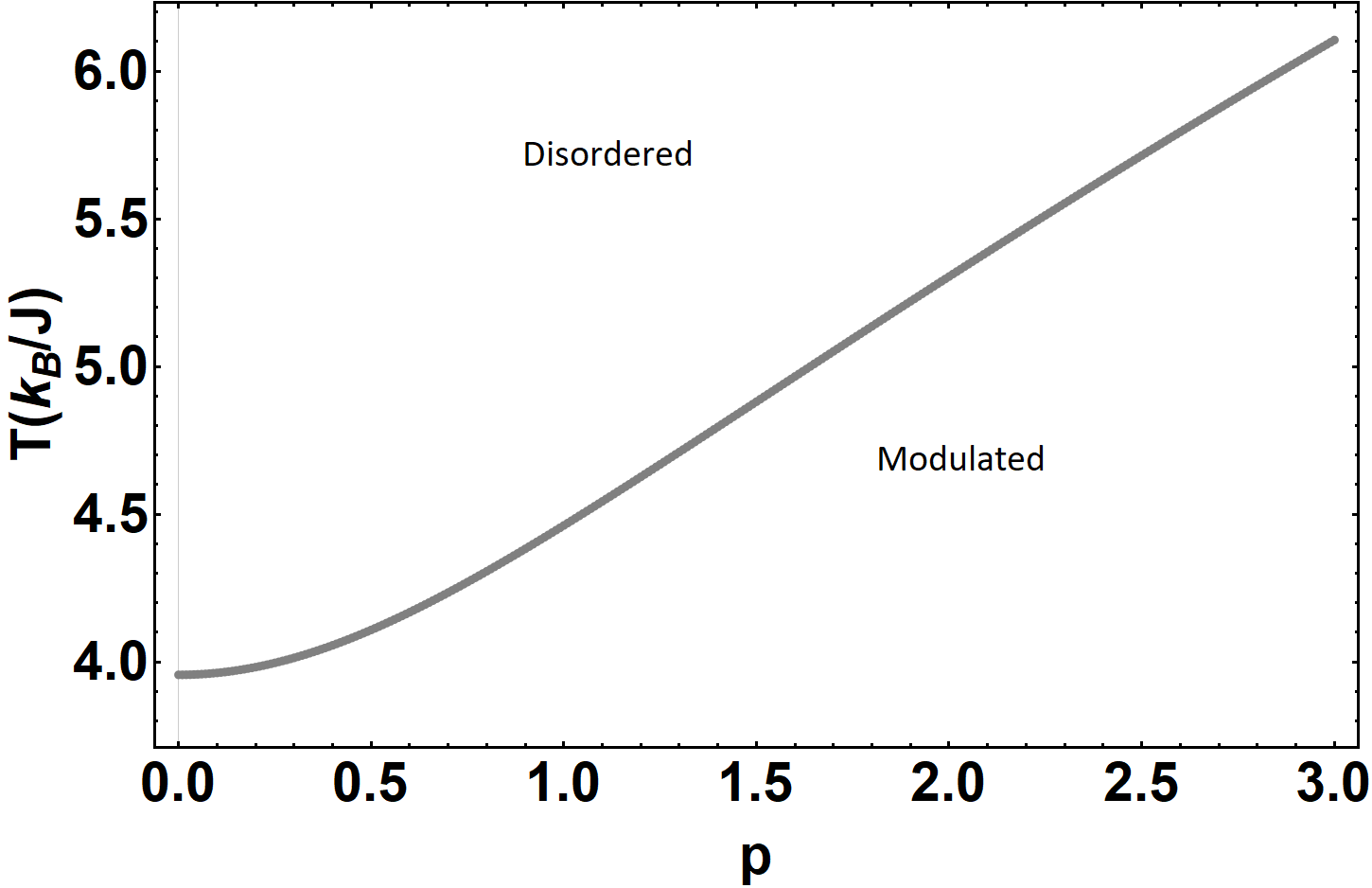}
    \caption{Transition border in the $T$ $-$ $p$ phase diagram.}
    \label{fig1:Ima1}
  \end{figure}

\section{Quantum spherical model with DM interactions}

In a quantum version of the classical mean-spherical model, the spin variable
$S_{\vec{r}}^{a}$ becomes a position operator at lattice site $\overrightarrow
{r}$, which is canonically conjugated to a momentum operator $P_{\vec{r}%
}^{\alpha}$. We then introduce a set of momentum operators, $P_{\vec{r}%
}^{\alpha}$, with $\alpha=x,y,z$, and write the commutation relations%
\begin{equation}
\lbrack S_{\vec{r}}^{a},S_{\vec{r}^{\prime}}^{\alpha^{\prime}}]=0,\qquad
\lbrack P_{\vec{r}}^{\alpha},P_{\vec{r}^{\prime}}^{\alpha^{\prime}}%
]=0,\qquad\lbrack S_{\vec{r}}^{\alpha},P_{\vec{r}^{\prime}}^{\alpha^{\prime}%
}]=i\delta_{\overrightarrow{r},\overrightarrow{r}^{\prime}}\delta
_{\alpha,\alpha^{\prime}},
\end{equation}
where $\alpha$ and $\alpha^{\prime}$ are the three Cartesian coordinates. With
the introduction of a kinetic energy term, we have the quantum Hamiltonian of
this system,
\begin{equation}
\mathcal{H}_{q}=\mathcal{H}_{1}+\mathcal{H}_{2}+\mathcal{H}_{3},
\end{equation}
with%
\begin{equation}
\mathcal{H}_{1}=\frac{1}{2}g\sum_{\vec{r}}\overrightarrow{P}_{\vec{r}}^{2}%
+\mu_{1}\sum_{\vec{r}}(S_{r}^{x})^{2}+\mu_{2}\sum_{\vec{r}}(S_{r}^{y})^{2}%
+\mu_{3}\sum_{\vec{r}}(S_{r}^{z})^{2},
\end{equation}%
\begin{equation}
\mathcal{H}_{2}=-J\sum_{\left(  \overline{r},\overrightarrow{r}^{\prime
}\right)  }\vec{S}_{\vec{r}}\cdot\vec{S}_{\vec{r}^{\prime}}%
\end{equation}
and%
\begin{equation}
\mathcal{H}_{3}=-D\sum_{\vec{r}}\left[  S_{\vec{r}}^{x}S_{\vec{r}+\widehat{z}%
}^{y}-S_{\vec{r}}^{y}S_{\vec{r}+\widehat{z}}^{x}\right]  ,
\end{equation}
where we adopt the same quantum parameter $g>0$ for the three directions, but
assume three distinct spherical potentials. Although we use the same notation
as in equation (\ref{hq}), spin and momenta are standard quantum operators.

We now introduce bosonic operators of creation, $(a_{\vec{r}}^{\alpha
})^{\dagger}$, and of annihilation, $a_{\vec{r}}^{\alpha}$, and change to a
language of second quantization. We then write%
\begin{equation}
S_{\vec{r}}^{\alpha}=\frac{1}{\sqrt{2}}\left(  \frac{g}{2\mu_{\alpha}}\right)
^{1/4}[a_{\vec{r}}^{\alpha}+(a_{\vec{r}}^{\alpha})^{\dagger}]
\end{equation}
and
\begin{equation}
P_{\vec{r}}^{\alpha}=-\frac{i}{\sqrt{2}}\left(  \frac{2\mu_{\alpha}}%
{g}\right)  ^{1/4}[a_{\vec{r}}^{\alpha}-(a_{\vec{r}}^{\alpha})^{\dagger}],
\end{equation}
for $\alpha=x,y,z$. In the next step towards the diagonalization of the
Hamiltonian, we adopt periodic boundary conditions, and use a Fourier
representation,%
\begin{equation}
a_{\vec{r}}^{\alpha}=\frac{1}{\sqrt{N}}\sum_{\vec{q}}\eta_{\vec{q}}^{\alpha
}\exp\left(  i\vec{q}.\vec{r}\right)  ,
\end{equation}
in which the sum is restricted to the first, symmetric, Brillouin zone, and
the new bosonic operators, $\left\{  \eta_{\overrightarrow{q}}^{\alpha
}\right\}  $, obey canonical commutation relations,%
\begin{equation}
\lbrack\eta_{\vec{q}}^{\alpha},\eta_{\vec{q}^{\prime}}^{\gamma}]=0,\qquad
\lbrack(\eta_{\vec{q}}^{\alpha})^{\dagger},(\eta_{\vec{q}^{\prime}}^{\gamma
})^{\dagger}]=0,\qquad\lbrack\eta_{\vec{q}}^{\alpha},(\eta_{\vec{q}^{\prime}%
}^{\gamma})^{\dagger}]=\delta_{\overrightarrow{q},\overrightarrow{q}^{\prime}%
}\delta_{\alpha,\gamma}.
\end{equation}

In the Fourier space, we finally have%
\begin{equation}
\mathcal{H}_{1}=%
{\displaystyle\sum\limits_{\alpha}}
\frac{1}{2}N\left(  g\mu_{\alpha}\right)  ^{1/2}+%
{\displaystyle\sum\limits_{\alpha}}
{\displaystyle\sum\limits_{\overrightarrow{q}\geq0}}
\left(  2g\mu_{\alpha}\right)  ^{1/2}\left[  \left(  \eta_{\overrightarrow{q}%
}^{\alpha}\right)  ^{\dagger}\eta_{\overrightarrow{q}}^{\alpha}+\left(
\eta_{-\overrightarrow{q}}^{\alpha}\right)  ^{\dagger}\eta_{-\overrightarrow
{q}}^{\alpha}\right]  ,
\end{equation}%
\[
\mathcal{H}_{2}=%
{\displaystyle\sum\limits_{\alpha}}
{\LARGE \{}-\frac{1}{2}\left(  \frac{g}{2\mu_{\alpha}}\right)  ^{1/2}%
{\displaystyle\sum\limits_{\overrightarrow{q}\geq0}}
\widehat{J}\left(  \overrightarrow{q}\right)  \left[  \eta_{\overrightarrow
{q}}^{\alpha}\eta_{-\overrightarrow{q}}^{\alpha}+\left(  \eta_{\overrightarrow
{q}}^{\alpha}\right)  ^{\dagger}\left(  \eta_{-\overrightarrow{q}}^{\alpha
}\right)  ^{\dagger}\right]  -
\]%
\begin{equation}
-\frac{1}{2}\left(  \frac{g}{2\mu_{\alpha}}\right)  ^{1/2}%
{\displaystyle\sum\limits_{\overrightarrow{q}\geq0}}
\widehat{J}\left(  \overrightarrow{q}\right)  \left[  \left(  \eta
_{\overrightarrow{q}}^{\alpha}\right)  ^{\dagger}\eta_{\overrightarrow{q}%
}^{\alpha}+\left(  \eta_{-\overrightarrow{q}}^{\alpha}\right)  ^{\dagger}%
\eta_{\overrightarrow{-q}}^{\alpha}\right]  {\LARGE \}},
\end{equation}
and%
\[
\mathcal{H}_{3}=-D\left(  \frac{g^{2}}{4\mu_{x}\mu_{y}}\right)  ^{1/4}%
{\displaystyle\sum\limits_{\overrightarrow{q}\geq0}}
i\,\sin q_{z}\,{\LARGE \{}\,\eta_{\overrightarrow{q}}^{x}\eta
_{-\overrightarrow{q}}^{y}-\left(  \eta_{\overrightarrow{q}}^{x}\right)
^{\dagger}\left(  \eta_{-\overrightarrow{q}}^{y}\right)  ^{\dagger}%
-\eta_{-\overrightarrow{q}}^{x}\eta_{\overrightarrow{q}}^{y}+
\]%
\begin{equation}
+\left(  \eta_{-\overrightarrow{q}}^{x}\right)  ^{\dagger}\left(
\eta_{\overrightarrow{q}}^{y}\right)  ^{\dagger}+\eta_{\overrightarrow{q}}%
^{x}\left(  \eta_{\overrightarrow{q}}^{y}\right)  ^{\dagger}-\left(
\eta_{\overrightarrow{q}}^{x}\right)  ^{\dagger}\eta_{\overrightarrow{q}}%
^{y}-\eta_{-\overrightarrow{q}}^{x}\left(  \eta_{-\overrightarrow{q}}%
^{y}\right)  ^{\dagger}+\left(  \eta_{-\overrightarrow{q}}^{x}\right)
^{\dagger}\eta_{-\overrightarrow{q}}^{y}\,{\LARGE \}},
\end{equation}
with the definition%
\begin{equation}
\widehat{J}\left(  \overrightarrow{q}\right)  =2J\left(  \cos q_{x}+\cos
q_{y}+\cos q_{z}\right)  .
\end{equation}

Taking into account the couplings between terms dependent on $+\overrightarrow
{q}$ and of $-\overrightarrow{q}$, we have to consider the quadratic form%
\[
Q=%
{\displaystyle\sum\limits_{\alpha}}
{\LARGE \{}\left(  2g\mu_{\alpha}\right)  ^{1/2}\left[  \left(  \eta
_{\overrightarrow{q}}^{\alpha}\right)  ^{\dagger}\eta_{\overrightarrow{q}%
}^{\alpha}+\left(  \eta_{-\overrightarrow{q}}^{\alpha}\right)  ^{\dagger}%
\eta_{-\overrightarrow{q}}^{\alpha}\right]  -
\]%
\[
-\left(  \frac{g}{2\mu_{\alpha}}\right)  ^{1/2}\frac{1}{2}\widehat{J}\left(
\overrightarrow{q}\right)  \left[  \eta_{\overrightarrow{q}}^{\alpha}%
\eta_{-\overrightarrow{q}}^{\alpha}+\left(  \eta_{\overrightarrow{q}}^{\alpha
}\right)  ^{\dagger}\left(  \eta_{-\overrightarrow{q}}^{\alpha}\right)
^{\dagger}+\left(  \eta_{\overrightarrow{q}}^{\alpha}\right)  ^{\dagger}%
\eta_{\overrightarrow{q}}^{\alpha}+\left(  \eta_{-\overrightarrow{q}}^{\alpha
}\right)  ^{\dagger}\eta_{-\overrightarrow{q}}^{\alpha}\right]  {\LARGE \}}-
\]%
\[
-D\left(  \frac{g^{2}}{4\mu_{x}\mu_{y}}\right)  ^{1/4}\left(  i\sin
q_{z}\right)  \,{\LARGE [}\,\eta_{\overrightarrow{q}}^{x}\eta
_{-\overrightarrow{q}}^{y}-\left(  \eta_{\overrightarrow{q}}^{x}\right)
^{\dagger}\left(  \eta_{-\overrightarrow{q}}^{y}\right)  ^{\dagger}%
-\eta_{-\overrightarrow{q}}^{x}\eta_{\overrightarrow{q}}^{y}+\left(
\eta_{-\overrightarrow{q}}^{x}\right)  ^{\dagger}\left(  \eta_{\overrightarrow
{q}}^{y}\right)  ^{\dagger}+
\]%
\begin{equation}
+\eta_{\overrightarrow{q}}^{x}\left(  \eta_{\overrightarrow{q}}^{y}\right)
^{\dagger}-\left(  \eta_{\overrightarrow{q}}^{x}\right)  ^{\dagger}%
\eta_{\overrightarrow{q}}^{y}-\eta_{-\overrightarrow{q}}^{x}\left(
\eta_{-\overrightarrow{q}}^{y}\right)  ^{\dagger}+\left(  \eta
_{-\overrightarrow{q}}^{x}\right)  ^{\dagger}\eta_{-\overrightarrow{q}}%
^{y}\,{\LARGE ]}\,.\label{QQ}%
\end{equation}

\subsection{Ferromagnetic sector}

The quadratic form (\ref{QQ}) has two different sectors, so that operators
associated with the $z$ direction can be treated separately. We then write
\begin{equation}
Q=Q_{z}+Q_{xy},
\end{equation}
with%
\begin{equation}
Q_{z}=A_{z}\left[  \left(  \eta_{\overrightarrow{q}}^{z}\right)  ^{\dagger
}\eta_{\overrightarrow{q}}^{z}+\left(  \eta_{-\overrightarrow{q}}^{z}\right)
^{\dagger}\eta_{-\overrightarrow{q}}^{z}\right]  +B_{z}\left[  \eta
_{\overrightarrow{q}}^{z}\eta_{-\overrightarrow{q}}^{z}+\left(  \eta
_{\overrightarrow{q}}^{z}\right)  ^{\dagger}\left(  \eta_{-\overrightarrow{q}%
}^{z}\right)  ^{\dagger}\right]  ,
\end{equation}
where%
\begin{equation}
A_{z}=\left(  2g\mu_{3}\right)  ^{1/2}\left[  1-\frac{1}{4\mu_{3}}\widehat
{J}\left(  \overrightarrow{q}\right)  \right]
\end{equation}
and%
\begin{equation}
B_{z}=-\left(  2g\mu_{3}\right)  ^{1/2}\frac{1}{4\mu_{3}}\widehat{J}\left(
\overrightarrow{q}\right)  .
\end{equation}

It is immediate to use a standard Bogoliubov transformation
\cite{Bogoliubov1982} to write $Q_{z}$ in a diagonal form in terms of a new
set of bosonic operators. According to this well-known procedure, and
discarding all the constant terms, we write%
\begin{equation}
Q_{z}=\lambda_{\overrightarrow{q}}\left[  \left(  \alpha_{\overrightarrow{q}%
}^{z}\right)  ^{\dagger}\alpha_{\overrightarrow{q}}^{z}+\left(  \alpha
_{-\overrightarrow{q}}^{z}\right)  ^{\dagger}\alpha_{-\overrightarrow{q}}%
^{z}\right]  ,
\end{equation}
where $\left\{  \alpha_{\overrightarrow{q}}^{z}\right\}  $ is a set of
transformed boson operators, associated with the energy spectrum%
\begin{equation}
\lambda_{\overrightarrow{q}}=\left\{  2g\left[  \mu_{3}-\frac{1}{2}\widehat
{J}\left(  \overrightarrow{q}\right)  \right]  \right\}  ^{1/2},\label{fspec}%
\end{equation}
which is the well known result for the quantum ferromagnetic mean-spherical
model \cite{Vojta1996} \cite{Oliveira2006}\cite{Bienzobaz2012}.

We now use the energy spectrum, given by equation (\ref{fspec}), and take care
of the proper constant terms of the spin Hamiltonian. We then have the
diagonal form%
\begin{equation}
\mathcal{H}_{z}=%
{\displaystyle\sum\limits_{\overrightarrow{q}}}
\lambda_{\overrightarrow{q}}\left[  \left(  \alpha_{\overrightarrow{q}}%
^{z}\right)  ^{\dagger}\alpha_{\overrightarrow{q}}^{z}+\frac{1}{2}\right]  ,
\end{equation}
from which we obtain the partition function
\begin{equation}
Z_{z}=%
{\displaystyle\prod\limits_{\overrightarrow{q}}}
\left\{
{\displaystyle\sum\limits_{n=0}^{\infty}}
\exp\left[  -\beta\lambda_{\overrightarrow{q}}\left(  n+\frac{1}{2}\right)
\right]  \right\}  .
\end{equation}
Taking into account the form of the spectrum of energy in this ferromagnetic
sector, we have%
\begin{equation}
N=-\frac{1}{\beta}\frac{\partial}{\partial\mu}\ln Z_{z}=%
{\displaystyle\sum\limits_{\overrightarrow{q}}}
\frac{g}{2\lambda_{\overrightarrow{q}}}\coth\left(  \frac{1}{2}\beta
\lambda_{\overrightarrow{q}}\right)  .
\end{equation}
In the classical limit, $g\rightarrow0$, we obtain a much simpler expression,%
\begin{equation}
N\rightarrow\frac{1}{2\beta}%
{\displaystyle\sum\limits_{\overrightarrow{q}}}
\frac{1}{\mu-\frac{1}{2}\widehat{J}\left(  \overrightarrow{q}\right)  },
\end{equation}
which can be shown to lead to the same equation (\ref{fborder}) for the
ferromagnetic border as we have obtained in the preceding section.

\subsection{Modulated sector}

We now sketch some calculations for the other sector of the quadratic form. As
in the classical case, there is a symmetry that relates the spherical
potential variables. We then make $\mu_{1}=\mu_{2}=\mu$. Also,we simplify the
notation to emphasize the couplings involving four bosonic modes, and discard
the vector symbols. We then write%
\[
Q_{xy}=A_{\bot}\,\left\{  a_{q}^{\dag}a_{q}+a_{-q}^{\dag}a_{-q}+b_{q}^{\dag
}b_{q}+b_{-q}^{\dag}b_{-q}\right\}  +
\]%
\[
+B_{\bot}\,\left\{  a_{q}a_{-q}+a_{q}^{\dag}a_{-q}^{\dag}+b_{q}b_{-q}%
+b_{q}^{\dag}b_{-q}^{\dag}\right\}  +
\]%
\begin{equation}
+\,C_{\bot}\,\{a_{q}b_{-q}-a_{q}^{\dag}b_{-q}^{\dag}-a_{-q}b_{q}+a_{-q}^{\dag
}b_{q}^{\dag}+a_{q}b_{q}^{\dagger}-a_{q}^{\dagger}b_{q}-a_{-q}b_{-q}^{\dagger
}+a_{-q}^{\dagger}b_{-q}\},
\end{equation}
with%
\begin{equation}
A_{\bot}=\left(  2g\mu\right)  ^{1/2}\left[  1-\frac{1}{4\mu}\widehat
{J}\left(  \overrightarrow{q}\right)  \right]  ,\label{Ap}%
\end{equation}%
\begin{equation}
B_{\bot}=-\left(  2g\mu\right)  ^{1/2}\frac{1}{4\mu}\widehat{J}\left(
\overrightarrow{q}\right)  ,\label{Bp}%
\end{equation}
and%
\begin{equation}
C_{\bot}=iD\left(  \frac{g}{2\mu}\right)  ^{1/2}\left(  \sin q_{z}\right)
.\label{Cp}%
\end{equation}

In order to analyze this four-component system, it is easier to write a set of
transformed bosonic operators,%
\begin{equation}
\gamma_{q}=wa_{q}+xb_{q}+ya_{-q}^{\dagger}+zb_{-q}^{\dag},
\end{equation}
so that%
\begin{equation}
\left[  \gamma_{q},Q_{xy}\right]  =E_{q}\,\gamma_{q},
\end{equation}
where $E_{q}$ is a bosonic energy spectrum \cite{Bogoliubov1982}
\cite{Hopfield1958}. We then write%
\begin{equation}
\left[  a_{q},Q_{xy}\right]  =A_{\bot}\,a_{q}+B_{\bot}\,a_{-q}^{\dag
}+\,C_{\bot}\,(-b_{q}-b_{-q}^{\dag}),
\end{equation}%
\begin{equation}
\left[  b_{q},Q_{xy}\right]  =A_{\bot}\,b_{q}+B_{\bot}\,b_{-q}^{\dag
}+\,C_{\bot}\,(a_{q}+a_{-q}^{\dag}),
\end{equation}%
\begin{equation}
\left[  a_{-q}^{\dag},Q_{xy}\right]  =-A_{\bot}\,a_{-q}^{\dag}-B_{\bot}%
\,a_{q}\,+C_{\bot}\,(b_{q}+b_{-q}^{\dag}),
\end{equation}
and%
\begin{equation}
\left[  b_{-q}^{\dag},Q_{xy}\right]  =-A_{\bot}\,b_{-k}^{\dag}-B_{\bot}%
\,b_{q}+\,C_{\bot}\,(-a_{q}-a_{-q}^{\dag}),
\end{equation}
from which we have%
\[
\left[  A_{\bot}w-B_{\bot}y+C_{\bot}(x-z)\right]  a_{q}+\left[  A_{\bot
}x-B_{\bot}z+C_{\bot}(y-w)\right]  b_{q}+
\]%
\[
+\left[  B_{\bot}w-A_{\bot}y+C_{\bot}(x-z)\right]  a_{-q}^{\dag}+
\]%
\begin{equation}
+\left[  B_{\bot}x-A_{\bot}z+C_{\bot}(y-w)\right]  b_{-q}^{\dag}=E_{q}\left[
wa_{q}+xb_{q}+ya_{-q}^{\dagger}+zb_{-q}^{\dag}\right]  .
\end{equation}
This equation lay be written in a matrix form,%
\begin{equation}
\mathbf{M}_{q}\,\left(
\begin{array}
[c]{c}%
w\\
x\\
y\\
z
\end{array}
\right)  =E_{q}\,\left(
\begin{array}
[c]{c}%
w\\
x\\
y\\
z
\end{array}
\right)  ,
\end{equation}
with%
\begin{equation}
\mathbf{M}_{q}=\left(
\begin{array}
[c]{cccc}%
A_{\bot} & C_{\bot} & -B_{\bot} & -C_{\bot}\\
-C_{\bot} & A_{\bot} & C_{\bot} & -B_{\bot}\\
B_{\bot} & C_{\bot} & -A_{\bot} & -C_{\bot}\\
-C_{\bot} & B_{\bot} & C_{\bot} & -A_{\bot}%
\end{array}
\right)  .
\end{equation}

From the eigenvalues of this matrix, we obtain the two branches of the
spectrum of energy of the diagonal bosonic system,%
\begin{equation}
\lambda^{2}=A_{\perp}^{2}-B_{\perp}^{2}\pm2\sqrt{C_{\perp}^{2}\left(
-A_{\perp}^{2}+2A_{\perp}B_{\perp}-B_{\perp}^{2}\right)  }.
\end{equation}
Taking into account that $C_{\perp}$ is a pure imaginary number, we can also
write%
\begin{equation}
\lambda^{2}=A_{\perp}^{2}-B_{\perp}^{2}\pm2\left\vert C_{\perp}\right\vert
\left(  A_{\perp}-B_{\perp}\right)  .
\end{equation}

We now take into account that $A_{\perp}$, $B_{\perp}$, and $C_{\perp}$, are
given by equations (\ref{Ap}), (\ref{Bp}), and (\ref{Cp}), and write the
expressions%
\begin{equation}
-C_{\perp}^{2}=D^{2}\frac{g}{2\mu}\sin^{2}q_{z},
\end{equation}%
\begin{equation}
A_{\perp}-B_{\perp}=\left(  2g\mu\right)  ^{1/2}\left[  1-\frac{1}{4\mu
}\widehat{J}\left(  \overrightarrow{q}\right)  \right]  +\left(  2g\mu\right)
^{1/2}\frac{1}{4\mu}\widehat{J}\left(  \overrightarrow{q}\right)  =\left(
2g\mu\right)  ^{1/2},
\end{equation}
and%
\begin{equation}
A_{\perp}^{2}-B_{\perp}^{2}=\left(  2g\mu\right)  \left[  1-\frac{1}{4\mu
}\widehat{J}\left(  \overrightarrow{q}\right)  \right]  ^{2}-\left(
2g\mu\right)  \frac{1}{16\mu^{2}}\left[  \widehat{J}\left(  \overrightarrow
{q}\right)  \right]  ^{2}=2g\mu-g\widehat{J}\left(  \overrightarrow{q}\right)
,
\end{equation}
in which the form of $C_{\perp}$ already gives an indication of the
oscillation along the axis. Using these expressions, we finally have the
energy spectrum,
\begin{equation}
\left[  \Lambda\left(  \overrightarrow{q}\right)  \right]  ^{2}=2g\left[
\mu-\frac{1}{2}\widehat{J}\left(  \overrightarrow{q}\right)  \pm D\sin
q_{z}\right]  ,\label{spec}%
\end{equation}
which is a characteristic result for an oscillating alignment along the $z$
axis. It is worth to point out the similarities with the energy spectrum of
ferromagnetic sector, given by equation (\ref{fspec}). Also, it is easy to
perform calculations to recover the classical spectrum.

With $D=0$, we recover the well-known quantum ferromagnetic case
\cite{Oliveira2006},
\begin{equation}
\left[  \Lambda_{Ferro}\left(  \overrightarrow{q}\right)  \right]
^{2}=2g\left[  \mu-\frac{1}{2}\widehat{J}\left(  \overrightarrow{q}\right)
\right]  .
\end{equation}

If we restrict the problem to a linear chain, along the $z$ direction,
equation (\ref{spec}) can be written as%
\begin{equation}
\left[  \Lambda_{chain}\left(  \overrightarrow{q}\right)  \right]
^{2}=2g\left[  \mu-J\cos q_{z}\pm D\sin q_{z}\right]  ,
\end{equation}
which indicates the oscillations along the $z$ direction (and which is similar
to the exact calculations for the energy spectrum of an $XY$ chain with DM
interactions \cite{Ming2013}).%

\begin{figure}[H]
    \centering
    \includegraphics[width=0.8\textwidth, height=0.5\textwidth]{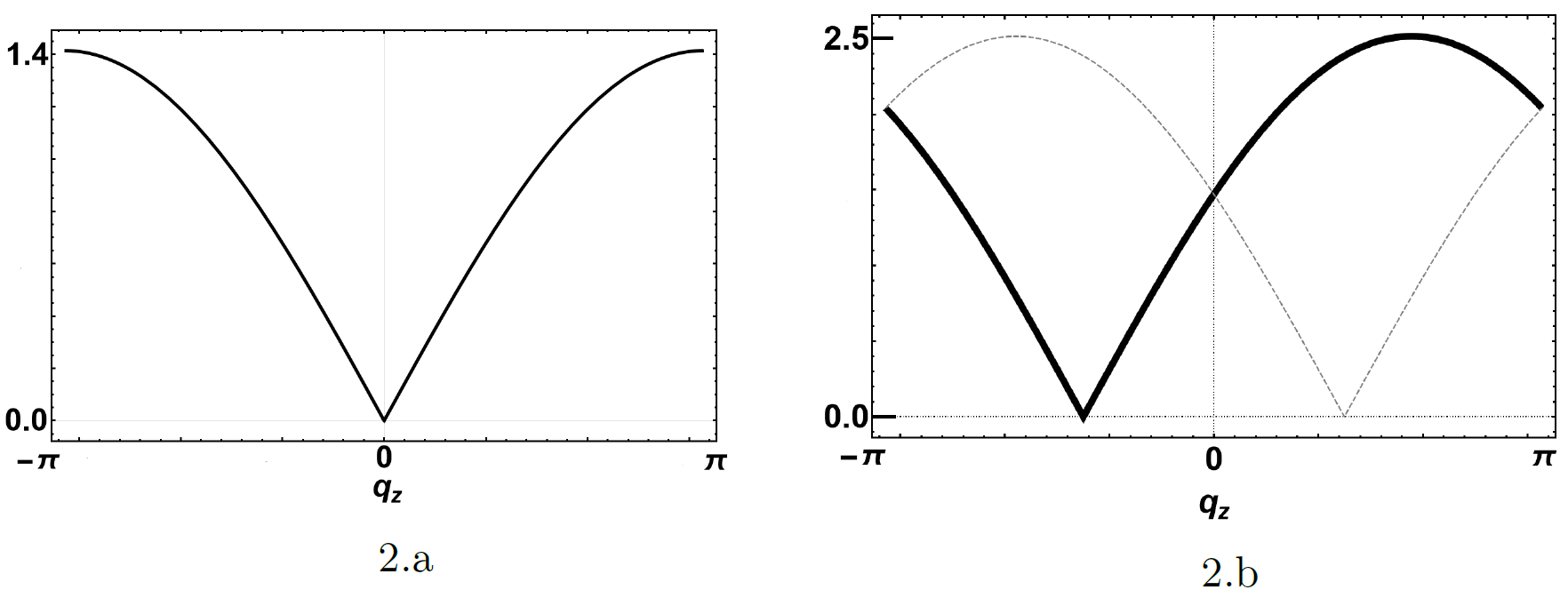}
    \caption{Figure (a) is a plot of the spectrum of energy in the ferromagnetic
sector. In figure (b), we plot the spectrum in the modulated sector (solid and
dashed lines represent positive and negative values of the ratio $D/J$).}
\label{duo}
    \label{fig2:Ima2}
  \end{figure}

In figure (2a) we plot the spectrum of energy in the ferromagnetic sector for
the maximun value of the spherical potential. In figure (2b), we show a
similar plot, for the modulated sector, with the maximum value of the
associated spherical potential. In this second figure, the minimum of energy
is shifted to $q_{z}\neq0$, depending on the sign of $d=D/J$.

\section{Conclusions}

We performed some exact calculations to investigate the phase transitions in a
ferromagnetic mean-spherical model, with the consideration of three distinct
spin components, and the addition of monoaxial Dzyaloshinkii-Moriya (DM) interactions.

In the classical case, we show the existence of a modulated structure along
the $\widehat{z}$\ direction at sufficiently low temperatures. We then define
a quantum version of this model system, which can be analyzed by standard
techniques of second quantization. We obtain the quantum energy spectrum, and
give arguments to show the persistence of spacial modulated structures in the
low-temperature region of the phase diagram.

\section{Acknowledgments}

We acknowledge conversations with professor Walter Wreszinski.

\end{document}